\documentclass[onecolumn]{revtex4}

\usepackage[utf8]{inputenc}
\usepackage{graphicx}
\usepackage{epsfig}
\usepackage{bm}
\usepackage{amsfonts}
\usepackage{amssymb}
\usepackage{color}
\usepackage{float}
\usepackage{amsmath}
\usepackage{enumitem}
\usepackage{wasysym}
\usepackage{dcolumn}
\usepackage{hyperref}
\usepackage{array}
\usepackage{lscape}
\usepackage{accents}
\usepackage{soul}
\usepackage{stmaryrd}
\usepackage{graphicx,subfigure,bm,color,psfrag,hyperref}
\usepackage{mathtools}
\usepackage{amsmath,amsthm}
\usepackage{verbatim}
\usepackage[dvipsnames]{xcolor}
\usepackage{refstyle}

\hypersetup{
    colorlinks=true,
    linkcolor=blue,
    filecolor=magenta,      
    citecolor=red
}

\newcommand{\udt}[3]{#1^{#2}_{\phantom{#2}#3}}
\newcommand{\udut}[4]{#1^{#2\phantom{#3}#4}_{\phantom{#2}#3\phantom{#4}}}

\newcommand{\dut}[3]{#1_{#2}^{\phantom{#2}#3}}
\newcommand{\dudt}[4]{#1_{#2\phantom{#3}#4}^{\phantom{#2}#3}}

\newcommand{\lc}[1]{\accentset{\circ}{#1}}

\begin{document}

\title{Cosmological perturbations in modified teleparallel gravity models: Boundary term extension
}

\author{Sebastian Bahamonde}
\email{sbahamonde@ut.ee, sebastian.beltran.14@ucl.ac.uk}
\affiliation{Laboratory of Theoretical Physics, Institute of Physics, University of Tartu, W. Ostwaldi 1, 50411 Tartu, Estonia}
\affiliation{Laboratory for Theoretical Cosmology, Tomsk State University of
Control Systems and Radioelectronics, 634050 Tomsk, Russia (TUSUR)}

\author{Viktor Gakis}
\email{vgakis@central.ntua.gr}
\affiliation{Institute of Space Sciences and Astronomy, University of Malta, Msida, Malta}
\affiliation{Department of Physics, National Technical University of Athens, Zografou Campus GR 157 73, Athens, Greece}

\author{Stella Kiorpelidi}
\email{skiorpel@central.ntua.gr}
\affiliation{Department of Physics, National Technical University of Athens, Zografou Campus GR 157 73, Athens, Greece}

\author{Tomi Koivisto}
\email{tomi.koivisto@ut.ee}
\affiliation{Laboratory of Theoretical Physics, Institute of Physics,University of Tartu, W. Ostwaldi 1, 50411 Tartu, Estonia}
\affiliation{National Institute of Chemical Physics and Biophysics, R\"avala pst. 10, 10143 Tallinn, Estonia}

\author{Jackson Levi Said}
\email{jackson.said@um.edu.mt}
\affiliation{Institute of Space Sciences and Astronomy, University of Malta, Msida, Malta}
\affiliation{Department of Physics, University of Malta, Msida, Malta}

\author{Emmanuel N. Saridakis}
\email{msaridak@phys.uoa.gr}
\affiliation{Department of Physics, National Technical University of Athens, Zografou Campus GR 157 73, Athens, Greece}
\affiliation{Department of Astronomy, School of Physical Sciences, University of Science and Technology of China, Hefei 230026, P.R. China}

\begin{abstract}
Teleparallel gravity offers a new avenue in which to construct gravitational models beyond general relativity. While teleparallel gravity can be framed in a way to be dynamically equivalent to general relativity, its modifications are mostly not equivalent to the traditional route to modified gravity. $f(T,B)$ gravity is one such gravitational theory where the second and fourth order contributions to the field equations are decoupled. In this work, we explore the all important cosmological perturbations of this new framework of gravity. We derive the gravitational propagation equation, its vector perturbation stability conditions, and its scalar perturbations. Together with the matter perturbations, we derive the effective gravitational constant in this framework, and find an interesting branching behaviour that depends on the particular gravitational models being probed. We close with a discussion on the relation of these results with other gravitational theories.
\end{abstract}
\maketitle

\section{Introduction}
Cosmological perturbations have shown the possibility of opening a pathway to revealing the cosmological evolution of the Universe in General Relativity (GR) and crucially in theories beyond GR \cite{Mukhanov:1990me}. The results of perturbations analysis can then be used in the confrontation with observational data to better understand which models fair better against data related to cosmic evolution \cite{Capozziello:2019cav,Nojiri:2017ncd,Nojiri:2010wj,Clifton:2011jh}. On the other hand, the $\Lambda$CDM cosmological model is supported by an abundance of evidence in describing the evolution of the Universe at all cosmological scales \cite{misner1973gravitation,dodelson2003modern} when matter beyond the standard model of particle physics is included. This takes the form of dark matter as a stabilizing ingredient in galactic structures \cite{Baudis:2016qwx,Bertone:2004pz}, while dark energy is represented by the cosmological constant \cite{Peebles:2002gy,Copeland:2006wr} and is the agent responsible for producing late-time accelerated cosmic expansion \cite{Riess:1998cb,Perlmutter:1998np} in this picture of the Universe. Nevertheless, even though great efforts have been directed at this part of the theory, internal problems persist with the concept of a cosmological constant \cite{Weinberg:1988cp}, and direct evidence for dark matter particles remains elusive \cite{Gaitskell:2004gd}.

The performance of the $\Lambda$CDM model has also become an open problem in recent years. In essence, the $\Lambda$CDM model was realised as a confrontation with Hubble expansion data but the so-called $H_0$ tension calls this feature into question, where the observational discrepancy between model independent measurements in the late Universe \cite{Riess:2019cxk,Wong:2019kwg} are in a meaning disagreement with the predicted value from the early Universe \cite{Aghanim:2018eyx,Ade:2015xua}. This tension has only grown in recent years \cite{Gomez-Valent:2019lny,Aghanim:2018eyx}. Saying that the problem still appears to be open with measurements from the tip of the red giant branch (TRGB, Carnegie-Chicago Hubble Program) pointing to a lower $H_0$ tension, the issue may ultimately be resolved by novel future observations such as measurements using gravitational wave astronomy standard candles \cite{Graef:2018fzu,Abbott:2017xzu} which may be accelerated once the LISA mission \cite{Baker:2019nia,Audley:2017drz} starts taking data.

There is now an abundance of theories beyond GR which aim to produce viable models of gravity that can agree with the new regime of precision measurements which have only became available in recent decades \cite{Clifton:2011jh,Capozziello:2011et,Barack:2018yly}. It is not enough for these theories to agree with cosmological observations at background level such as with the value of $H_0$. Theories beyond GR must also produce observable quantities from their perturbed dynamical equation that agree with current observations to be seriously considered. One such observable that is gaining increased interest is that of $f\sigma_8$ which also hosts a growing but weak tension with the $\Lambda$CDM model of cosmology. It was in Refs.\cite{Bardeen:1980kt,Kodama:1985bj} that cosmological perturbation theory was first developed in a consistent way, where a gauge-invariant approach was first developed. This approach has been used to analyze numerous models of gravity \cite{Clifton:2011jh} with various successes. These theories mainly appear as an extension to GR \cite{Sotiriou:2008rp,Faraoni:2008mf,Capozziello:2011et} and build on corrections designed for various purposes that may have a cosmological effect at different epochs. However, these approaches can be collectively grouped by their common expression of gravitation through the use of the Levi-Civita connection, i.e. they communicate gravity by means of geometric curvature of spacetime \cite{misner1973gravitation,nakahara2003geometry}. This is not the only choice where torsion, through teleparallel gravity, has become an increasingly popular replacement for the curvature associated with the Levi-Civita connection \cite{Aldrovandi:2013wha,Cai:2015emx,Krssak:2018ywd}.

Teleparallel Gravity (TG) refers to the collection of theories that express gravity through the torsion of the teleparallel connection \cite{Weitzenbock1923}. The general linear teleparallel connection \cite{Jimenez:2019ghw} is only required to be flat (curvature-less), but in this work we further restrict to the case of metric-compatible teleparallel connections. Given these properties, all curvature based measures of gravity will naturally vanish identically. A consequence of this is that the Einstein-Hilbert action, as determined with the teleparallel connection, will also vanish, i.e., $R=0$, while its regular Levi-Civita connection version will remain the same, i.e., $\lc{R}\neq 0$ (where over-circles will refer to quantities determined using the Levi-Civita connection throughout). By replacing the Ricci scalar in the Einstein-Hilbert action with its torsion scalar analog will produce identical dynamical equations. This is called the {Teleparallel equivalent of General Relativity} (TEGR), and differs from GR by a boundary $B$ term in the gravitational Lagrangian. 

The TEGR boundary term embodies the fourth order contributions to the field equations which is an important aspect of many theories beyond GR. In TG, the second and fourth order field equation contributions become decoupled from each other unlike in standard gravity where the Levi-Civita connection is employed. Using this rationale, modifications of TEGR will have a meaningful and impactful difference as compared with regular modified theories of gravity. The most prescient of these properties will be the realisation of producing generically second order theories of gravity in some generalizations of TEGR. This is to be contrasted with GR where by the Lovelock theorem \cite{Lovelock:1971yv}, second order field equations are only produced by the Einstein-Hilbert action (with the addition of a constant) unless extra assumptions are included such as scalar fields or extra dimensions. In TG, the Lovelock theorem is weakened \cite{Gonzalez:2015sha,Bahamonde:2019shr} allowing for a plethora of additional theories beyond TEGR that continue to produce second order field equations. TG also has a number of other attractive properties such as its similarity to Yang-mills theory \cite{Aldrovandi:2013wha} which gives it features of particle physics theory, as well as the possibility of giving a well-defined energy-momentum tensor for gravitation \cite{Blixt:2018znp,Blixt:2019mkt}, and that it does not require an associated Gibbons--Hawking--York boundary term giving a more structured form to its Hamiltonian formalism, in addition to others.

One of the best studied modification to GR is that of $f(\lc{R})$ gravity \cite{Sotiriou:2008rp,Faraoni:2008mf,Capozziello:2011et}, and TEGR can similarly be generalized to produce $f(T)$ gravity \cite{Ferraro:2006jd,Ferraro:2008ey,Bengochea:2008gz,Linder:2010py,Chen:2010va,Bahamonde:2019zea}. This has several key properties chief among which is that it produced second order field equations and has shown promise in its confrontation with observations at various scales \cite{Cai:2015emx,Nesseris:2013jea,Farrugia:2016qqe,Finch:2018gkh,Farrugia:2016xcw,Iorio:2012cm,Ruggiero:2015oka,Deng:2018ncg}. TG can also offer a path in which $f(\lc{R})$ gravity is dynamically generalized by considering $f(T,B)$ gravity where the different order contributions are decoupled from one another \cite{Bahamonde:2015zma,Capozziello:2018qcp,Bahamonde:2016grb,Farrugia:2018gyz,Bahamonde:2016cul,Wright:2016ayu}. This limits to $f(\lc{R})$ gravity in the limit where $f(\lc{R})=f(-T+B)$. $f(T,B)$ gravity has shown promise as being a viable model at various scales ranging from solar system tests in the weak field regime \cite{Farrugia:2020fcu,Capozziello:2019msc,Farrugia:2018gyz,Bahamonde:2020bbc}, as well as its cosmological theoretical structure \cite{Bahamonde:2015zma,Bahamonde:2016grb,Bahamonde:2016cul,Bahamonde:2015zma} and confrontation with observational data \cite{Escamilla-Rivera:2019ulu}. In terms of solar system tests, Ref.\cite{Bahamonde:2020bbc} explores the predictions for perihelion shift, deflection of light, the Cassini experiment, Shapiro delay and gravitational redshift predictions for an $f(T,B)$ model that contains a combination of power law contributions in terms of the two scalar $T$ and $B$. The analysis is performed for a leading order perturbative metric about a Schwarzschild background. The ensuing predictions are compared with solar system observations with the result being that the $f(T,B)$ model under consideration is compatible with these observations. This work also explores other effects such as the black hole photon sphere and the impact of $f(T,B)$ gravity on the equivalence principle. Another important work on the topic is Ref.\cite{Farrugia:2020fcu} which investigates the gravitomagnetic effects for a perturbed $f(T,B)$ Lagrangian. In this case, predictions for the Lens-Thirring and geodetic effects are compared with the available observations from Gravity Probe B \cite{Everitt:2011hp}. In this instance, again consistent model parameter values are determined. In the cosmological regime, in addition to the numerous foundational works on the topic \cite{Capozziello:2019msc,Farrugia:2018gyz,Bahamonde:2015zma,Bahamonde:2016grb,Bahamonde:2016cul,Caruana:2020szx}, the analysis contained in Ref.\cite{Escamilla-Rivera:2019ulu} probes a number of literature models against recent Hubble parameter data with a number of models showing promising results.

In $f(T)$ gravity, cosmological perturbations have been considered in a number of works \cite{Zheng:2010am,Anagnostopoulos:2019miu,Nunes:2018evm,Nesseris:2013jea,Farrugia:2016qqe,DAgostino:2018ngy} which has been extended to a number of other extensions to TEGR such as Ref.\cite{Farrugia:2016pjh} where matter perturbations are considered in $f(T,\mathcal{T})$ gravity and Ref.\cite{Haro:2013bea} in which the perturbations in teleparallel loop quantum cosmology are performed. In the present work, we determine the cosmological perturbations about a flat Friedmann–Lema\^{i}tre–Robertson–Walker (FLRW) metric. Together with the perturbations associated with the matter contribution, we form the linear perturbation equations in order to produce probes that can be used in observational cosmology. In section~\ref{f_T_B_review}, we briefly review $f(T,B)$ gravity and its associated cosmology. In section~\ref{f_T_B_cosmic_pert}, we develop the gravitational perturbations while in section~\ref{matter_pert} we form the perturbations equations with the perturbations about a perfect fluid. Finally in section~\ref{conclu}, we conclude our work with a discussion of the core results. In this work we use the $(+,-,-,-)$ signature.

\section{Modified teleparallel theories of gravity\label{f_T_B_review}}

The curvature associated with the Levi-Civita connection $\mathring{\Gamma}^{\sigma}_{\mu\nu}$ (we use over-circles to denote quantities calculated with the Levi-Civita connection throughout) is torsion-less and satisfies the metricity condition \cite{nakahara2003geometry,misner1973gravitation}. TG is distinct from GR in that it supplants this connection with a torsion-ful teleparallel connection $\Gamma^{\sigma}_{\mu\nu}$ that has vanishing curvature and continues to satisfy the metricity condition \cite{ortin2004gravity,Aldrovandi:2013wha,Cai:2015emx,Krssak:2018ywd}. In GR, many quantities are built on the Riemann tensor since it gives a measure of curvature on a manifold, it is for this reason that many modified theories of gravity feature implementations of this tensor \cite{Clifton:2011jh}. However, in replacing this connection with its flat counterpart in teleparallel gravity, renders these quantities null irrelevant of the entries of the metric tensor. It is in this context that TG theory requires a novel approach to constructing tensorial quantities in order to build gravitational models.

GR is built on the metric tensor $g_{\mu\nu}$ being the fundamental dynamical object, as are the modifications of GR. In TG, the metric tensor becomes a derived object with the tetrad $\udt{e}{A}{\mu}$ replacing it as the fundamental gravitational variable of the theory \cite{Aldrovandi:2013wha}. In this context, Latin indices refer to the Minkowski space while Greek indices point to the general manifold, where the tetrad acts as a soldering agent between the two. Thus, the tetrad (and its inverses $\dut{E}{A}{\mu}$) can transform between the general manifold and its associated Minkowski space through
\begin{eqnarray}\label{metric_tetrad_eq}
    g_{\mu\nu}=\udt{e}{A}{\mu}\udt{e}{B}{\nu}\eta_{AB}\,,& &\eta_{AB}=\dut{E}{A}{\mu}\dut{E}{B}{\nu}g_{\mu\nu}\,,
\end{eqnarray}
where the tetrads observe orthogonality conditions
\begin{eqnarray}
    \udt{e}{A}{\mu}\dut{E}{B}{\mu}=\delta^B_A\,,& &\udt{e}{A}{\mu}\dut{E}{A}{\nu}=\delta^{\nu}_{\mu}\,,
\end{eqnarray}
for consistency's sake. The teleparallel connection can then be defined as \cite{Weitzenbock1923}
\begin{equation}
    \Gamma^{\sigma}_{\nu\mu} := \dut{E}{A}{\sigma}\partial_{\mu}\udt{e}{A}{\nu} + \dut{E}{A}{\sigma}\udt{\omega}{A}{B\mu}\udt{e}{B}{\nu}\,,
\end{equation}
where $\udt{\omega}{A}{B\mu}$ represents the spin connection. The teleparallel connection represents the most general linear affine connection that is flat and satisfies the metricity condition \cite{Aldrovandi:2013wha,Hehl:1994ue}. The spin connection $\udt{\omega}{A}{B\mu}$ acts as a balance to retain the general covariance of the ensuing field equations due to the freedom in the choice of the tetrad components in Eq.~(\ref{metric_tetrad_eq}) \cite{Krssak:2015oua}. Levi-Civita based theories (such as GR) hide this feature in its inertial structure and does not play an active role for most expressions of the theory \cite{misner1973gravitation,nakahara2003geometry}. The spin connection in TG is totally inertial and incorporates the effects of the local Lorentz transformations (LLTs) thus producing LLT invariant theories. Naturally, there will always exist a frame where the spin connection is vanishing as in the original formulation in Ref.\cite{Weitzenbock1923}, and this choice of frame is called the Weitzenb\"{o}ck gauge (WG).

The spin connection can be fully represented as $\udt{\omega}{A}{B\mu}=\udt{\Lambda}{A}{C}\partial_{\mu}\dut{\Lambda}{B}{C}$ \cite{Aldrovandi:2013wha}, where the full breadth of the LLTs (Lorentz boosts and rotations) are represented by $\udt{\Lambda}{A}{B}$. Through this perspective, there exist an infinite number of tetrads that satisfy Eq.~(\ref{metric_tetrad_eq}), each of which produces an independent spin connection which counter-balances each other. It is therefore the tetrads together with its associated spin connection that renders a covariant formulation of TG.

Building on rationale of the Riemann tensor, the teleparallel connection can be straightforwardly used to build a meaningful measure of torsion through an antisymmetric operation on its lower indices. Thus, torsion can be represented as an expression of antisymmetry through the torsion tensor defined as \cite{Krssak:2018ywd,Cai:2015emx}
\begin{equation}
    \udt{T}{\sigma}{\mu\nu} :=- 2\Gamma^{\sigma}_{\left[\mu\nu\right]}\,,
\end{equation}
where square brackets denote the usual antisymmetric operator. The field strength of TG is represented by the torsion tensor \cite{Aldrovandi:2013wha}, which transforms covariantly under both diffeomorphisms and LLTs. As in theories of gravity based on the Levi-Civita connection, we can also construct other gravitational tensors that reveal general features of TG. Firstly, take the contorsion tensor that emerges as the difference between the teleparallel and Levi-Civita connections, and can be written purely in terms of the torsion tensor as 
\begin{equation}
    \udt{K}{\sigma}{\mu\nu} := \Gamma^{\sigma}_{\mu\nu} - \mathring{\Gamma}^{\sigma}_{\mu\nu} = \frac{1}{2}\left(\dudt{T}{\mu}{\sigma}{\nu} + \dudt{T}{\nu}{\sigma}{\mu} - \udt{T}{\sigma}{\mu\nu}\right)\,.
\end{equation}
This has an important role to play in relating TG with GR and its modifications, as will become apparent later on. Another core component of TG is the superpotential defined as \cite{Krssak:2018ywd}
\begin{equation}
    \dut{S}{A}{\mu\nu} := \frac{1}{2}\left(\udt{K}{\mu\nu}{A} + \dut{e}{A}{\nu}T^{\mu} - \dut{e}{A}{\mu}T^{\nu}\right)\,,
\end{equation}
where $T^{\nu}:=\udut{T}{\alpha}{\alpha}{\nu}=-\udt{T}{\alpha\nu}{\alpha}$. This has been shown to have a potential relationship to the energy-momentum tensor for gravitation \cite{Aldrovandi:2004db} but the issue remains open \cite{Koivisto:2019jra}. By contracting the torsion tensor together with its superpotential, the torsion scale emerges \cite{Cai:2015emx}
\begin{equation}\label{torsion_scalar_def}
    T := \dut{S}{A}{\mu\nu}\udt{T}{A}{\mu\nu}\,,
\end{equation}
as being purely the product of the teleparallel connection, in an analogous way to the Ricci scalars dependence purely on the Levi-Civita connection. The standard Ricci scalar $\lc{R}$ (computed with the Levi-Civita connection) clearly will not vanish but its TG analog will, $R=0$. Using the contorsion tensor, it can be shown that the teleparallel Ricci scalar, which vanishes, is equal to the sum of the Ricci and torsion scalars (up to a boundary term) through \cite{Hayashi:1979qx,Hehl:1976kj}
\begin{equation}
    R=\lc{R} + T - \frac{2}{e}\partial_{\mu}\left(e T^{\mu}\right) = 0\,.
\end{equation}
This directly leads to an equivalency relation between the standard Ricci and torsion scalars given by
\begin{equation}\label{Ricci_torsion_equiv}
    \lc{R} = -T + \frac{2}{e}\partial_{\mu}\left(e T^{\mu}\right) = -T + 2\mathring{\nabla}_{\mu}\left(T^{\mu}\right) = -T + B\,,
\end{equation}
where we define the boundary term as
\begin{equation}
    B:=2\mathring{\nabla}_{\mu}\left(T^{\mu}\right)\,,
\end{equation}
called the TEGR boundary term, and where $e = \text{det}\left(\udt{e}{A}{\mu}\right)=\sqrt{-g}$ is the tetrad determinant. The ensuing dynamical equations will thus be guaranteed to be identical since these scalars differ by a boundary when expressed linearly. In this way, we can define the 
Teleparallel Gravity equivalent of general relativity (TEGR) as
\begin{equation}
    \mathcal{S}_{\text{TEGR}} = -\frac{1}{2\kappa^2}\int d^4 x\, eT + \int d^4 x\, e\mathcal{L}_m\,,
\end{equation}
where $\kappa^2=8\pi G$ and $\mathcal{L}_m$ is the regular matter Lagrangian. The boundary term difference at the level of the Lagrangians can have an important impact when modifications of TEGR are considered which can lead to novel approaches to gravity that not recoverable in GR. In fact, the boundary term embodies the fourth order derivative contributions to the GR field equations thus decoupling these contributions that are incorporated in the Ricci scalar in standard gravity.

In standard gravity, one of the most popular approaches to gravity beyond GR is that of $f(\lc{R})$ gravity \cite{Sotiriou:2008rp,Capozziello:2011et} in which the Ricci scalar is straightforwardly generalized to an arbitrary function therefore. Another is Horndeski theory in which a single scalar field is added with the proviso of producing second order equations of motion \cite{Horndeski:1974wa} which was recently formulated in TG \cite{Bahamonde:2019shr,Bahamonde:2019ipm,Bahamonde:2020cfv}. In TEGR, two scalars play an important role in producing the equivalency with GR in standard gravity. The torsion scalar produces the same second order dynamics, while the boundary term absorbs the divergence quantities. The $T$ and $B$ scalars embody the second and fourth order contributions respectively. It is for these reasons that to fully embody the rationale of many theories beyond GR we must consider an arbitrary generalization with both scalars. This will also suitably incorporate $f(\lc{R})$ gravity as a subcase of the broader $f(T,B)$ framework.

$f(T,B)$ gravity \cite{Bahamonde:2015zma,Capozziello:2018qcp,Bahamonde:2016grb,Paliathanasis:2017flf,Farrugia:2018gyz,Bahamonde:2016cul,Bahamonde:2016cul,Wright:2016ayu} is a novel approach to modifying gravity and limits to $f(\lc{R})$ gravity in the limits where $f(T,B)=f(-T+B)=f(\lc{R})$. This is expressed as a generalization of TEGR through the action
\begin{equation}\label{f_T_action}
    \mathcal{S}_{f(T,B)} = \frac{1}{2\kappa^2}\int d^4 x\, ef(T,B) + \int d^4 x\, e\mathcal{L}_m\,.
\end{equation}

Taking a variation of the action with respect to the tetrad gives \cite{Bahamonde:2015zma,Farrugia:2018gyz}
\begin{eqnarray}
    2\delta_{\nu}^{\lambda}\mathring{\Box}f_{B}-2\mathring{\nabla}^{\lambda}\mathring{\nabla}_{\nu}f_{B}+Bf_{B}\delta_{\nu}^{\lambda}+4\Big[(\partial_{\mu}f_{B})+(\partial_{\mu}f_{T})\Big]S_{\nu}{}^{\mu\lambda}& &\nonumber \\
    +4e^{-1}e^{A}{}_{\nu}\partial_{\mu}(eS_{A}{}^{\mu\lambda})f_{T}-4f_{T}T^{\sigma}{}_{\mu\nu}S_{\sigma}{}^{\lambda\mu}-f\delta_{\nu}^{\lambda} & =& 2\kappa^{2}\udt{\Theta}{\nu}{\lambda}\,,\label{field_equations}
\end{eqnarray}
where subscripts denote derivatives, and $\udt{\Theta}{\nu}{\lambda}$ is the regular energy-momentum tensor for matter. The dynamical equations here have been derived within the WG that has been shown to be compatible with a flat FLRW metric \cite{Bahamonde:2015zma,Capozziello:2018qcp,Bahamonde:2016grb,Farrugia:2018gyz,Caruana:2020szx} which is what we develop here.
One could argue that the field equations do not assume a manifestly Lorentz covariant form. We remind the reader that the action of the Lorentz group is associated with tetrad - spin connection pairs i.e, they both need to be transformed in the same time. Since, the field equations are derived directly in the WG and the spin connection is trivialised, loss of the manifest Lorentz covariance is to be expected. Nevertheless, the field equations in the WG are still Lorentz covariant just not manifestly. This can be directly proven by starting from the WG and then re-introducing the spin connection via an arbitrary Lorentz transformation of the form
\begin{equation}\label{eq:tetrad_spin_con_trans}
e^A{}_{\mu} \mapsto e'^A{}_{\mu} = \Lambda^A{}_Be^B{}_{\mu}\,,
\end{equation}
\begin{equation}
\omega'^A{}_{B\mu} = \Lambda^A{}_C\partial_{\mu}(\Lambda^{-1})^C{}_B\,.
\end{equation}
Henceforth, the use of WG is at the cost of the manifest character of the Lorentz covariance but rather not the Lorentz covariance itself. This kind of misunderstanding has led to a false idea of ``broken Lorentz invariance`` in the past, regarding (modified) teleparallel theories in general.

The spectrum of $f(T,B)$ gravity in Minkowski spacetime ~\cite{Koivisto:2018loq} includes the usual massless graviton with a $\sim -f_T$ modulation of the propagator, and an additional “scalaron” with a mass $\sim 1/\sqrt{-f_{BB}}$. Thus, to avoid a ghost one requires that $f_T<0$ and to avoid a tachyon that $f_{BB}<0$. A feature of general $f(T,B)$ gravity thus is that it expresses the same gravitational wave polarization signature as $f(\lc{R})$ gravity \cite{Farrugia:2018gyz,Capozziello:2018qcp}. The result of Ref. ~\cite{Koivisto:2018loq} also suggests that when the self-interaction of the extra scalar degree of freedom can be neglected, its presence leads to disagreement with the Solar system tests of gravity theory, in analogy to the $f(R)$ models which predict value $\gamma=1/2$ for the Eddington parameter $\gamma$ which is tightly constrained by the experimental data to its value $\gamma=1$ in General Relativity\footnote{The post-Newtonian analysis of teleparallel modified gravity by Flathmann and Hohmann also reproduces the prediction $\gamma=1/2$ (assuming a massless field and restricting to the case $-T+f(B)$ which is covered by their analysis) \cite{Flathmann:2019khc}.} Thus, as is the case in $f(R)$ gravity, in general $f(T,B)$ gravity models need to implement some kind of screening mechanism in order to comply with the local tests of gravity.

The field equations in Eq.~(\ref{field_equations}) can straightforwardly be rewritten as
\begin{equation}
   -f_{T}\lc{G}_{\mu\nu}+g_{\mu\nu}\lc{\Box} f_{B}-\lc{\nabla}_{\mu}\lc{\nabla}_{\nu}f_{B} +
  \frac{1}{2}(Bf_{B}+Tf_{T}-f)g_{\mu\nu}
  +2\Big[\lc{\nabla}^{\lambda}f_T+\lc{\nabla}^{\lambda}f_B\Big]S_{\nu\lambda\mu}= \kappa^2 \Theta_{\mu\nu} \,,
  \label{FieldEq2}
\end{equation}
where the Einstein tensor $\lc{G}_{\mu\nu}$ explicitly emerges due to the close relationship between curvature and torsion. It is important to point out that while this represents the field equations of the teleparallel $f(T,B)$ gravity, the Einstein tensor and the covariant derivatives are dependent on the Levi-Civita connection. It is useful to separate these equations to its symmetric and antisymmetric parts. To do this, let us introduce the following tensor
\begin{equation}
    Q_{\nu\mu}:=\frac{1}{2}\Big[\lc{\nabla}^{\lambda}f_T+\lc{\nabla}^{\lambda}f_B\Big]S_{\nu\lambda\mu}\,,
\end{equation}
and then the antisymmetric part of the field equations Eq.~(\ref{FieldEq2}) becomes
\begin{eqnarray}
Q_{[\alpha\nu]}&=&(\partial^\lambda f_{T}+\partial^\lambda f_{B})(T_{\alpha\lambda\nu}+g_{\alpha\lambda}T_{\nu}-g_{\alpha\nu}T_{\lambda})\\
&=&\Big[(f_{TT}+f_{TB})\partial^{\lambda}T+(f_{BB}+f_{TB})\partial^{\lambda}B\Big](T_{\alpha\lambda\nu}+g_{\alpha\lambda}T_{\nu}-g_{\alpha\nu}T_{\lambda})=0\,,
\label{anti_fe}
\end{eqnarray}
where we have used the condition that the energy-momentum tensor is symmetric. It is very important, for the consistency of the theory, that the antisymmetric part presented in Eq.(\ref{anti_fe}) is always satisfied properly, in order to attain consistency in the use of WG.

\noindent Now, in this work we probe the cosmology of $f(T,B)$ gravity through the tetrad
\begin{equation}\label{flrw_tetrad}
    \udt{e}{A}{\mu}=\textrm{diag}(1,a(t),a(t),a(t))\,,
\end{equation}
where $a(t)$ is the scale factor, and which reproduces the flat homogeneous isotropic FLRW metric
\begin{equation}
    ds^2=dt^2-a(t)^2(dx^2+dy^2+dz^2)\,,
\end{equation}
through Eq.~(\ref{metric_tetrad_eq}). This diagonal tetrad is compatible with a flat spin connection, $\udt{\omega}{A}{B\mu}=0$ \cite{Krssak:2015oua,Tamanini:2012hg}. Through Eq.~(\ref{torsion_scalar_def}), the torsion scalar turns out to be
\begin{equation}
    T =-6H^2\,,
\end{equation}
and the boundary term is given by 
\begin{equation}
    B =-6(3H^2+\dot{H})\,,
\end{equation}
which together reproduce the Ricci scalar, i.e. $\lc{R}=-T+B = -6(\dot{H} + 2H^2)$. Using the field equations in Eq.~(\ref{field_equations}) together with the FLRW tetrad in Eq.~(\ref{flrw_tetrad}) produces the Friedmann equations
\begin{eqnarray}
    3H(\dot{f}_{B}-2Hf_{T})+\frac{1}{2}(Bf_{B}-f) &=& \kappa^{2}\rho\,,\\
    -\ddot{f}_{B}+2f_{T}\dot{H}+2H(3Hf_{T}+\dot{f}_{T})+\frac{1}{2}(f - Bf_{B}) &=& \kappa^{2}P\,,
\end{eqnarray}
where overdots refer to derivatives with respect to cosmic time $t$, and where $\rho$ and $P$ respectively represent the energy density and pressure of matter.

At background level, we can write the Friedmann equations for $f(T,B)$ gravity as an effective fluid equation as an addition to the TEGR Lagrangian through $f(T,B)\rightarrow-T + \tilde{f}(T,B)$. Evaluating the dynamical equations in Eq.~(\ref{field_equations}) for the FLRW setting gives the Friedmann equations
\begin{eqnarray}
    3H^2 &=& \kappa^2 \left(\rho+\rho_{\text{eff}}\right)\,,\\
    3H^2 + 2\dot{H} &=& -\kappa^2\left(P+P_{\text{eff}}\right)\,.
\end{eqnarray}
Through the effective fluid description, this means that the fluid properties are represented by
\begin{eqnarray}
    \kappa^2 \rho_{\text{eff}} &:=& 3H^2\left(3\tilde{f}_B + 2\tilde{f}_T\right) - 3H\dot{\tilde{f}}_B + 3\dot{H}\tilde{f}_B + \frac{1}{2}\tilde{f}\,, \label{eq:friedmann_mod}\\
    \kappa^2 P_{\text{eff}} &:=& -\frac{1}{2}\tilde{f}-\left(3H^2+\dot{H}\right)\left(3\tilde{f}_B + 2\tilde{f}_T\right)-2H\dot{\tilde{f}}_T+\ddot{\tilde{f}}_B\,.
\end{eqnarray}
The $\tilde{f}(T,B)$ gravity effective fluid description also satisfies the fluid equation \cite{Bahamonde:2016grb}
\begin{equation}
    \dot{\rho}_{\text{eff}}+3H\left(\rho_{\text{eff}}+P_{\text{eff}}\right) = 0\,,
\end{equation}
and leads directly to an effective fluid equation of state (EoS)
\begin{eqnarray}\label{EoS_func}
    \omega_{\text{eff}} &:=& \frac{P_{\text{eff}}}{\rho_{\text{eff}}}\\ & = & -1+\frac{\ddot{\tilde{f}}_B-3H\dot{\tilde{f}}_B-2\dot{H}\tilde{f}_T-2H\dot{\tilde{f}}_T}{3H^2\left(3\tilde{f}_B+2\tilde{f}_T\right)-3H\dot{\tilde{f}}_B+3\dot{H}\tilde{f}_B-\frac{1}{2}\tilde{f}}\,. 
\end{eqnarray}
In the $\Lambda$CDM limit, this EoS approaches an effective cosmological constant behaviour where $\omega_{\text{eff}}=-1$, as expected. In the next section we consider the cosmological perturbations within $\tilde{f}(T,B)$ gravity. In that context, it is more convenient to work with a pure $f(T,B)$ gravity representation.

\section{Cosmological perturbations of \texorpdfstring{$f(T,B)$}{f(T,B)} gravity\label{f_T_B_cosmic_pert}}

Cosmological perturbations can reveal an incredible amount of information about the Universe that is not immediately clear from the background cosmology such as the formation of cosmic structures and the gravitational wave background universe. Cosmic perturbations were investigated in $f(T)$ gravity several times such as Ref.\cite{Chen:2010va} where the tetrad is only in the correct WG in terms of the tensor perturbations and thus results in an overly restrictive set of scalar perturbations, which is later clarified in Ref.\cite{Krssak:2015oua}. It was only in Ref.\cite{Zheng:2010am} that the situation was fully resolved, which was also applied to the $f(T,\mathcal{T})$ gravity scenario in Ref.\cite{Farrugia:2016pjh}. The core results have since been confirmed and widened in Refs.\cite{Izumi:2012qj,Wu:2012hs,Golovnev:2018wbh}. In what follows, we explore the tensor and scalar cosmological perturbations within the sub-horizon limit. This is achieved by taking the scalar-vector-tensor (SVT) decomposition of the cosmological perturbations using \cite{Golovnev:2018wbh}
\begin{equation}
    \left[\delta e^{A}{}_{\mu}\right]:=\left[\begin{array}{cc} \varphi & a\left(\partial_{i}\beta+\beta_{i}\right)\\
    \delta^{I}{}_{i}\left(\partial^{i}b+b^{i}\right) & a\delta^{Ii}\left(-\psi\delta_{ij}+\partial_{i}\partial_{j}h+2\partial_{(i}h_{j)}+\frac{1}{2}h_{ij}+\epsilon_{ijk}\left(\partial^{k}\sigma+\sigma^{k}\right)\right)
    \end{array}\right]\,,\label{eq:tetrad perturbation-1}
\end{equation}
which inherits its symmetries from the metric and retains the WG even at perturbative level. It is important to emphasize that this tetrad remains a good tetrad even at perturbative level in that the associated spin connection components are compatible with the case where they vanish. This is crucial to producing a consistent cosmological perturbation analysis. A note on the use of indices, A,B,C,D,.. and Greek lowercase letters $\mu,\nu,\rho,\sigma,..$ are used as 4-D indices on the Minkowski and general manifold respectively. The middle range Latin indices I,J,K,.. and i,i,k,.. refer to spacial 3-D indices in Minkowski and general manifold respectively. In fact, this produces the regular perturbed metric 
\begin{equation}
\left[\delta g_{\mu\nu}\right]=\left[\begin{array}{cc}
-2\varphi & a\left(\partial_{i}(b-\beta)+\left(b_{i}-\beta_{i}\right)\right)\\
a\left(\partial_{i}\left(b-\beta\right)+\left(b_{i}-\beta_{i}\right)\right) & 2a^{2}\left(-\psi\delta_{ij}+\partial_{i}\partial_{j}h+2\partial_{(i}h_{j)}+\frac{1}{2}h_{ij}\right)
\end{array}\right]\,,\label{eq:metric perturbation}
\end{equation}
due to Eq.~(\ref{metric_tetrad_eq}), where $h_{ij}$ is symmetric, traceless $h_{ij}\delta^{ij}=0$, and transverse $\partial^{i}h_{ij}=0$, while all the vectors are solinoidal $\partial_{i}b^{i}=0$. 

Now, in our convention, the Fourier transform of a perturbation $X$ will be given by
\begin{equation}
    X\left(t,x\right)=\int\frac{d^{3}k}{\left(2\pi\right)^{2/3}}\left[X(t,k)e^{ikx}+X^{\dagger}(t,k)e^{-ikx}\right]\,,
\end{equation}
which is used throughout to transform the cosmological perturbations. In the appendices we include all important calculations of each perturbation.

Also, the matter perturbation of the energy-momentum tensor for a perfect fluid $\delta \Theta_{\mu\nu}$ is
\begin{equation}
    \delta \Theta_{\mu\nu}=(\delta \rho +\delta P)u_\mu u_\nu+(\rho+P)\delta u_\mu u_\nu +(\rho+P)u_\mu \delta u_\nu + \delta P g_{\mu\nu}+P\delta g_{\mu\nu}\,,
\end{equation}
where the 4-velocity is represented by $u_\mu$ and 3-velocity by $v_i=\partial_i v$ with components
\begin{eqnarray}
    \delta \Theta_{00} &=& \delta \rho+2\rho\phi\,,\\
    \delta \Theta_{0i} &=& \delta \Theta_{i0}=-a \rho\partial_i(b-\beta)-a (\rho+P)\partial_i v\,,\\
    \delta \Theta_{ij} &=& a^2\delta P\delta_{ij}-2a^2P\psi\delta_{ij}+2a^2P\partial_i\partial_j h\,.
\end{eqnarray}
Together, this forms the basis for the matter perturbation equations to be explored later on after the scalar perturbations. In the following computations the  xAct packages \cite{Martin-Garcia:2007bqa,MartinGarcia:2008qz,DBLP:journals/corr/abs-0803-0862,Brizuela:2008ra,GomezLobo:2011xv,Pitrou:2013hga,Nutma:2013zea} were also used.

\subsection{Tensor Perturbations \label{subsec:Tensor-Perturbations}}

Considering the tensor perturbation part of the cosmological perturbations in the tetrad in Eq.~(\ref{eq:tetrad perturbation-1}) which are
\begin{equation}\label{ten_pert}
    \delta e^{I}{}_{j} = \frac{a}{2}\delta^{Ii}h_{ij}\,,
\end{equation}
we can determine the tensor perturbations within the $f(T,B)$ action. The tensor modes are determined by considering perturbations up to second order in the Lagrangian density, which in Fourier space results in the gravitational wave propagation equation
\begin{equation}
    \ddot{h}_{ij}+(3+\nu)H\dot{h}_{ij}+\frac{k^{2}}{a^{2}}h_{ij} = 0\,,\label{eq:GWPE}
\end{equation}
which governs propagation of tensor perturbations. The background equations were used in these derivations to simplify the perturbation results. Here, the Planck mass run rate turns out to be
\begin{equation}\label{ten_pert_stab}
    \nu=\frac{1}{H}\frac{\dot{f}_{T}}{f_{T}}\,,
\end{equation}
which is a frictional term in the propagation of gravitational waves, as evidenced through the gravitational wave propagation equation \cite{Ezquiaga:2017ekz,Saltas:2014dha,Riazuelo:2000fc}. Immediately, a stability condition in which $f_{T}<0$ can be read off (which depends on the convention being used for the torsion scalar). Another crucial point is that the Eq.~(\ref{ten_pert_stab}) describes a massless spin 2 field that propagates with the speed of light \cite{Copeland:2018yuh}.
Thus the propagation of gravitational waves is in total agreement with the multimessenger events of GW170817 \cite{TheLIGOScientific:2017qsa} and GRB170817A \cite{Goldstein:2017mmi}. Note that, exactly the same result holds true in the $f(T,B)\rightarrow f(T)$ limit as seen from \cite{Izumi:2012qj,Golovnev:2018wbh,Cai:2018rzd}. In addition, in the $f(T,B)\rightarrow f(-T+B)=f(\lc{R})$ limit where $f_T\rightarrow-f_R$,  Eq.~(\ref{eq:GWPE}) reproduces exactly the usual result found in the literature \cite{DeFelice:2010aj} where in this case $\nu=\dot{f}_{R}/Hf_{R}$.

In this context, $f(T,B)$ gravity is not strongly constrained by present observations since it predicts speed of light propagation of gravitational waves and no constraints exist for the Planck run rate. However, the stability conditions in Eq.~(\ref{ten_pert_stab}) will be crucial to forming stable models and have an impactful effect on the other perturbations that follow.

\subsection{Vector (and pseudovector) Perturbations}

The vector perturbations in the cosmological perturbations in Eq.~(\ref{eq:tetrad perturbation-1}) are represented by
\begin{eqnarray}
    \left[\delta e^{A}{}_{\mu}\right]= & \left[\begin{array}{cc}
    0 & a\beta_{i}\\
    \delta^{I}{}_{i}b^{i} & a\delta^{Ii}\epsilon_{ijk}\sigma^{k}
\end{array}\right]\,,\label{=0003B4evector}
\end{eqnarray}
where the gauge freedom is fixed by the choice $h_{i}\equiv0$. Using the field equations, we directly obtain the perturbation equations for the $\beta_{i}$ and the pseudovector $\sigma_{i}$
\begin{eqnarray}
    W_{[0i]}:\quad0&= & \text{\ensuremath{\sigma}}_{i}(\dot{f}_{B}+\dot{f}_{T})\,,\label{W=00005B0i=00005Dvector}\\
    W_{[ij]}(i\neq j):\quad0 &= & \beta_{i}(\dot{f}_{B}+\dot{f}_{T})\,,\label{W=00005Bij=00005Dvector}
\end{eqnarray}
which for $\dot{f}_{B}+\dot{f}_{T}\neq0$ give $\text{\ensuremath{\sigma}}_{i}=0$
and $\beta_{i}=0$. We are left with two equations that govern the evolution of $b_{i}$ and $v_{i}$ which are embodied through
\begin{eqnarray}
    W_{ij}(i\neq j):\quad0 &=& \dot{b}_{j}+b_{j}\left(2H+\frac{\dot{f}_{T}}{f_{T}}\right)\,,\label{Wijvector}\\
    W_{0i}:\quad0 &=& b_{i}\left(a(H(6\dot{f}_{B}+4\dot{f}_{T})+4f_{T}\dot{H}-2\ddot{f}_{B})-\frac{k^{2}f_{T}}{a}\right)\nonumber \\
    && \qquad+av_{i}(6H(\dot{f}_{B}-2Hf_{T}) + Bf_{B} - f)\,,\label{W0ivector}
\end{eqnarray}
which involves only those two components, and where $v_i$ represents the 3-velocity (discussed further in the appendix). Immediately, it is clear that if this is solved for $b_{i}$, then it is solvable for $v_{i}$ as well. At this stage one can directly see that the vector perturbations are not propagating since Eq.~(\ref{Wijvector}) is just a constraint equation and can further read off the stability condition $f_T<0$ which is exactly the same condition found in Subsection.\ref{subsec:Tensor-Perturbations} . Another important observation is that , which has exactly the same form as that reported in Ref.~\cite{Golovnev:2018wbh} for $f(T)$ gravity with the exception that in our case $f(T)\rightarrow f(T,B)$ (and the impact of this on derivative terms).

If  $\dot{f}_{B}+\dot{f}_{T}\equiv0$ it implies that $f_{T}=-f_{B}=-f_{R}$ which is the case of $f(\lc{R})$ gravity, where all antisymmetric field
equations are trivialised with $W_{[\mu\nu]}\equiv0$. By introducing $Y_{i}:=b_{i}-\beta_{i}$, we end up with the following nonvanishing field equations 
\begin{eqnarray}
    W_{ij}(i\neq j):\quad0 &=& \dot{Y}_{j}+Y_{j}\left(2H+\frac{\dot{f}_{R}}{f_{R}}\right)\,,\label{WijvectorfR}\\
    W_{0i}:\quad0 &=& av_{i}\left(6H(2Hf_{R}+\dot{f}_{R}) + Bf_{R} - f\right)\nonumber \\
    && \qquad+Y_{i}\left(2a(-2f_{R}\dot{H}+H\dot{f}_{R}-\ddot{f}_{R})+\frac{k^{2}f_{R}}{a}\right)\,,\label{W0ivectorfR}
\end{eqnarray}
where $f_{R}=df/d\lc{R}$. In this equation we notice that, again, there are not propagating vector perturbations which is a well known result in $f(\lc{R})$ theories \cite{DeFelice:2010aj}.

\subsection{Scalar Perturbations\label{eq:tetrad perturbation}}

Selecting the scalar perturbations of Eq.~(\ref{eq:tetrad perturbation-1}) gives the following linear perturbations
\begin{eqnarray}
\left[\delta e^{A}{}_{\mu}\right] & = & \left[\begin{array}{cc}
\varphi & a\partial_{i}\beta\\
\delta^{I}{}_{i}\partial^{i}b & a\delta^{Ii}\left(-\psi\delta_{ij}+\partial_{i}\partial_{j}h+\epsilon_{ijk}\partial^{k}\sigma\right)
\end{array}\right]\,,
\end{eqnarray}
in which we will adopt the Newtonian gauge where $b=\beta$ and $h=0$. In the following we report the final field equations but in the appendix, the component calculations that build up to these results are presented. The symmetric field equations of the scalar perturbations are given by
\begin{eqnarray}
W_{00}: \quad \kappa^{2}\delta\rho &=& 3H\delta\dot{f}_{B}+\Big(\frac{k^{2}}{a^{2}} + \frac{B}{2}\Big)\delta f_{B}-6H^{2}\delta f_{T} - \frac{1}{2}f_{T}\delta T-\frac{2Hk^{2}f_{T}}{a}b\label{FE_W00}\nonumber \\
 & & +\dot{\psi}(12Hf_{T}-3\dot{f}_{B})+\frac{2k^{2}f_{T}}{a^{2}}\psi+6H\phi(2Hf_{T}-\dot{f}_{B})\,,\\
\nonumber \\
W_{ij}(i\neq j): \quad \psi-\phi &=& \frac{1}{f_{T}}(a(\dot{f_{T}}+\dot{f}_{B})b-\delta f_{B})\label{Wij}\,,\\
\nonumber \\
W_{i}^{i}: \quad -\kappa^{2}\delta P &=& \delta\ddot{f}_{B}+\delta f_{B}\left(\frac{2k^{2}}{3a^{2}} + \frac{B}{2}\right)-2H\delta\dot{f}_{T}-2(3H^{2}+\dot{H})\delta f_{T} - \frac{1}{2}f_{T}\delta T\nonumber \\
 & & +2f_{T}\ddot{\psi}+2\dot{\psi}(6Hf_{T}+\dot{f}_{T})+\frac{2k^{2}f_{T}}{3a^{2}}\psi-\frac{2k^{2}}{3a}(\dot{f}_{B}+3Hf_{T}+\dot{f}_{T})b\nonumber \\
 & & +\dot{\phi}(2Hf_{T}-\dot{f}_{B})+\phi\left(4f_{T}\left(-\frac{2k^{2}f_{T}}{3a^{2}}+3H^{2}+\dot{H}-2\ddot{f}_{B}\right)+4H\dot{f}_{T}\right)\,,
\label{Wii}
\end{eqnarray}
where $\delta f_T = f_{TT}\delta T+f_{TB}\delta B$ and $\delta f_B = f_{BT}\delta T+f_{BB}\delta B$, while the antisymmetric contributions are
\begin{eqnarray}
W_{0i} \quad \kappa^{2}av(P+\rho) &=& \delta\dot{f}_{B}-3H\delta f_{B}+2f_{T}\dot{\psi}-2H\delta f_{T}+(2f_{T}H-\dot{f}_{B})\phi\,,\\
\nonumber \\
W_{i0}: \quad \kappa^{2}av(P+\rho) &=& \delta\dot{f}_{B}-H\delta f_{B}+2f_{T}\dot{\psi}+2(\dot{f}_{T}+\dot{f}_{B})\psi+(2f_{T}H-\dot{f}_{B})\phi\,,\\
\nonumber \\
W_{i0}-W_{0i}: \quad 0 &=& H(\delta f_{T}+\delta f_{B})+\psi(\dot{f_{T}}+\dot{f}_{B})\,,\label{anti}
\end{eqnarray}
and where the energy-momentum conservation in the case of dust (for the general case see the Appendix \ref{scalarappendix}) is given by
\begin{eqnarray}
\nabla_{\mu}\Theta_{0}{}^{\mu}: \quad \delta\dot{\rho}+3H\delta\rho &=& \frac{\rho}{a}k^{2}v+3\dot{\psi}\rho\,,\\
\nonumber \\
\nabla_{\mu}\Theta_{i}{}^{\mu}: \quad a\dot{v}+aHv &=& -\phi\,.\label{W[0i]}
\end{eqnarray}

These equations reduce to the usual ones found in the literature \cite{Zheng:2010am,Izumi:2012qj,Golovnev:2018wbh} for $f(T)$ gravity in the limit $f(T,B)\rightarrow f(T)$ where $f_B\rightarrow0$ and the antisymmetric part survives as expected in Eq.(\ref{anti}). In the limit of $f(T,B)\rightarrow f(-T+B)=f(\lc{R})$ where ($f_T\rightarrow -f_R,f_B\rightarrow f_R$), one can recover after a few trivial substitutions the usual equations \cite{DeFelice:2010aj}. In addition, in this limit the antisymmetric part of the field equations in Eq.(\ref{anti}) is trivialised and the scalar $b$ completely drops off from the field equations, as expected.

The scalar perturbations are coupled with the perturbations of the energy-momentum components and so this is not enough information to determine the impact of these cosmological perturbations on observational parameters. In the next section, we will study the matter perturbation equations  to determine the role of $f(T,B)$ gravity on the growth of structure in the Universe.

\section{Matter density perturbation equation in \texorpdfstring{$f(T,B)$}{f(T,B)} gravity\label{matter_pert}}

In this section, we consider dust for the perfect fluid, and derive the corresponding matter perturbation equations. Following Refs.~\cite{DeFelice:2010aj,Tsujikawa:2007gd}, we introduce the variable $V:=av$ and start by defining the density contrast $\delta_{m}$ as
\begin{equation}
\delta_{m}:=\frac{\delta\rho}{\rho}+3HV\label{delta_m_def}\,.
\end{equation}

In order to determine the time derivative of this parameter, we need to utilize the continuity equation to obtain the density parameter time derivative, which is
\begin{equation}
\delta \dot{\rho}+3 H \delta \rho =\frac{k^2 \rho V}{a^2}+3 \rho\dot{\psi}\,.\label{eq:Continuity EQ}
\end{equation}
The time derivative of the density contrast parameter can then be written as
\begin{eqnarray}
\dot{\delta}_{m} &=& -\frac{\nabla^{2}V}{a^{2}}+3\dot \psi+3\dot{(HV)}\,,\label{eq:=0003B4mdot-1}\\
\dot{V} &=& -\phi\,,\label{eq:Vdot-1}
\end{eqnarray}
where the time derivative of $V$ is also presented. By combining both derivatives, we obtain
\begin{equation}\label{matter_pert_equation}
\ddot{\delta}_{m}+2H\dot{\delta}_{m}=\frac{\nabla^{2}\phi}{a^{2}}+3\ddot \psi+3\ddot{(HV)}+6H\dot{\psi}+6H\dot{(HV)}\,.
\end{equation}

In the sub-horizon approximation $k>>aH$, $k$ being well inside the Hubble radius, the dominant terms are $k$ and $\delta\rho$. Now that we have all the prerequisites we need to proceed, let us first summarize the dominant terms in this limit
\begin{eqnarray}
\label{spacederaprox}
\left\{ \frac{k^{2}}{a^{2}}|\phi|,\frac{k^{2}}{a^{2}}|\psi|,\frac{k^{2}}{a^{2}}|\beta|,\frac{k^{2}}{a^{2}}|\delta f_{T}|,\frac{k^{2}}{a^{2}}|\delta f_{B}|\right\} & \gg\left\{ H^{2}|\phi|,H^{2}|\psi|,H^{2}|\beta|,H^{2}|\delta f_{T}|,H^{2}|\delta f_{B}|\right\} 
\end{eqnarray}
and
\begin{eqnarray}
\label{timederaprox}\dot{|X}| & \apprle{|HX|} & {\textstyle {\rm where}}\quad X\in\left\{ \phi,\psi,\beta,\delta f_{T},\delta f_{B},\dot{\phi},\dot{\psi},\dot{\beta},\delta\dot{f_{T}},\delta\dot{f_{B}}\right\}\,.
\end{eqnarray}
Thus, it follows directly that in Fourier space of the sub-horizon limit of Eq.~(\ref{matter_pert_equation})
\begin{equation}
    \ddot{\delta}_{m}+2H\dot{\delta}_{m} \simeq-\frac{k^{2}\phi}{a^{2}}=4\pi\rho G_{\rm eff}\delta_{\rm m} =\frac{\kappa^2}{2}\rho G_{\rm eff}\delta_{\rm m}\,,\label{meszaroseq}
\end{equation}
from which it follows that the only contributing scalar is $\phi$. Along a similar vein, $\Sigma_{\rm def}$ is a parameter sensitive to weak lensing which appears when we write the lensing potential $-\left(\phi+\psi\right)$ in terms of the matter density contrast $\delta_{m}$, so $\Sigma_{\rm def}$ plays a similar role to $G_{\rm eff}$ but between the lensing potential and $\delta_{m}$ specifically. This parameter is defined as 
\begin{equation}
    \Sigma :=\frac{1}{2}\frac{G_{\rm eff}}{G}\left(1+\frac{\psi}{\phi}\right)\,,\label{Sigmadef}
\end{equation}
which we will also calculate in conjunction with $G_{\rm eff}$ in what follows. We start from the sub-horizon approximation of the field equations in Eqs.~(\ref{FE_W00},\ref{W[0i]})
\begin{eqnarray}
    W_{00}: \quad \kappa^{2}\delta\rho &\simeq& \left(\frac{2k^{2}f_{T}}{a^{2}}-3H\dot{f}_{B}\right)\psi+\left(\frac{k^{2}}{a^{2}}-3\dot{H}\right)\delta f_{B}+6H(Hf_{T}-\dot{f}_{B})\phi-6H^{2}\delta f_{T}\,,\label{W00finalsubaprox}\\
    W_{[0i]}: \quad 0 &\simeq & \psi(\dot{f}_{B}+\dot{f}_{T})+H\delta f_{B}+H\delta f_{T}\,,\label{W=00005B0i=00005Dfinalsubaprox}\\
    W_{ij}(i\neq j): \quad 0 &=& -ab(\dot{f}_{B}+\dot{f}_{T})+\delta f_{B}+f_{T}\psi-f_{T}\phi\,,\label{Wijfinalsubaprox}\\
    W_{i}^{i}: \quad 0 &\simeq& \delta f_{B}(18a^{2}\dot{H}-4k^{2})+12a^{2}(4H^{2}+\dot{H})\ensuremath{\delta}f_{T}+4ak^{2}(\dot{f}_{B}+\dot{f}_{T})b\nonumber \\
    & & +\left(6a^{2}\left(H(\dot{f}_{B}-4\dot{f}_{T})+2\ddot{f}_{B}\right)+4f_{T}(k^{2}-6a^{2}\dot{H})\right)\phi\nonumber \\
    & & -4\psi\left(f_{T}k^{2}+3a^{2}H\dot{f}_{T}\right)\,,\label{Wiifinalsubaprox}
\end{eqnarray}
from which we present the fully expanded form of the $W_{[0i]}$ component 
\begin{eqnarray}
    W_{[0i]}: \quad\quad 0 &\simeq & -\frac{4H^{2}k^{2}(f_{BB}+2f_{TB}+f_{TT})}{a}b\nonumber \\
    & & +\frac{\left(a^{2}((\dot{f}_{B}+\dot{f}_{T})+12H^{3}(f_{TT}+f_{TB}))+4Hk^{2}(f_{BB}+f_{TB})\right)}{a^{2}}\psi\\
    & & -\frac{2H\left((f_{BB}+f_{TB})(k^{2}-6a^{2}\dot{H})-6a^{2}H^{2}(f_{TT}+f_{TB})\right)}{a^{2}}\phi\,.\label{W=00005B0i=00005Dfinalsubaprox EXPANDED}
\end{eqnarray}
Note that $W_{[0i]}$ is actually a constraint equation and so must be used in the solution process. Consequentially in order to have
a closed system we only need one more equation from $\left\{ W_{ij},W_{i}^{i}\right\} $,
which we choose to be $W_{ij}$. Henceforth our system will be comprised of $\left\{ W_{00},W_{[0i]},W_{ij}\right\}$. We checked in every case that the fourth equation $W_{i}^{i}$ was always satisfied. Before proceeding we define the useful parameters
\begin{align}
\Pi & :=f_{B}+f_{T}\,,\label{PI}\\
\Upsilon & :=f_{BB}+2f_{TB}+f_{TT}=\Pi_{T}+\Pi_{B}\,,\label{YPSILON}\\
\Xi & :=f_{TB}^{2}-f_{TT}f_{BB}=-\Pi_{T}\Pi_{B}+f_{TB}\Upsilon\,.\label{KSI}
\end{align}
One could think of $\Pi$ as the deviation from $f(\lc{R})$ gravity where $\Pi|_{f(\lc{R})}\equiv0$. These quantities will help us classify the $f(T,B)$ models in three branches 
\begin{enumerate}
\item $\left\{ \Pi\neq \text{const},\Upsilon\neq0\right\} $ \\
Which can further be classified using $\Xi=-\Pi_{T}\Pi_{B}+f_{TB}\Upsilon$
\begin{enumerate}
\item $\left\{ \Pi\neq\text{const},\Upsilon\neq0,\Xi\neq0\right\} $ most general case
of $f(T,B)$
\item $\left\{ \Pi\neq\text{const},\Upsilon\neq0,\Xi=0\right\} $ includes $f(T)$
\end{enumerate}
\item $\left\{ \Pi\neq\text{const},\Upsilon\equiv0\right\} $ \\
Which can further be classified using $\Upsilon\equiv0\Rightarrow\Pi_{B}\equiv-\Pi_{T}$
into Eq.~(\ref{KSI}) as $\Xi=\Pi_{T}^{2}=\Pi_{B}^{2}$ 
\begin{enumerate}
\item $\left\{ \Pi\neq\text{const},\Upsilon=0,\Xi\neq0\right\} $
\item $\left\{ \Pi = {\rm const},\Upsilon=0,\Xi=0\right\} $ the unique $f(\lc{R})$ case
\end{enumerate}
\end{enumerate}
The above branches may also be indicators of variable degrees of freedom (dof), since we know for sure that $f(\lc{R})$ has 3 dof. We also know that $f(T)$ ``varies'' in between 3-5 maximum dof \cite{Blagojevic:2020dyq,Li:2011rn,Ong:2013qja,Ferraro:2018tpu}.

We will elaborate a bit on the two major conditions $\Xi\equiv0$ and $\Upsilon\equiv0$. Starting off with $\Xi\equiv0$, it can be solved using separation of variables if one assumes $f(T,B)=f_{1}(T)f_{2}(B)$, then one finds
\begin{eqnarray}
    f(T,B) & = & f_{0}\Big((B+Bm-C_{2})(T+mT-C_{3}m)^{m}\Big)^{\frac{1}{m+1}}\,,\quad m\neq-1\,,\label{XIsol1}\\
    f(T,B) & = & f_{0}e^{C_{1}T+C_{2}B}\,,\quad m=-1\,.\label{XIsol2}
\end{eqnarray}
where $f_{0},C_{1},C_{2},C_{3},m$ are constants. Another family of solutions are of the form $f(T,B)=f(\Phi)$ where $\Phi=\Phi(T,B)$ i.e single variable dependence. Popular models of this type are where $\Phi\equiv\lc{R}=-T+B$ and $\Phi\equiv T$ which represent $f(\lc{R})$ and $f(T)$ theories of gravity respectively. Another form of single variable dependence is $f(TB)=c\sqrt{TB}$ which is the only acceptable model of the form $f(TB)=c\left(TB\right)^{m}$. Finally, a less known model of importance here is
\begin{equation}
    f(T,B)=-T+F(B)\,,\label{f(T,B)=00003D-=0003B3T+f(B)}
\end{equation}
which  will be used   later on on the analysis.

As for the condition $\Upsilon\equiv0$, it is satisfied by a family of solutions of the form 
\begin{equation}
    f(T,B)=f_{1}(\lc{R})X+f_{2}(\lc{R})\,,\label{solutionY}
\end{equation}
where $X=X(T,B)$ is any function such that $X_{T}+X_{B}\neq0$ and $\Upsilon\equiv0$. The condition $X_{T}+X_{B}\neq0$ practically means that $X\neq X(\lc{R})$ so that the total solution in Eq.~(\ref{solutionY}) is not reduced to just $f(\lc{R})$. The most intuitive form would be $X=\left(c_{1}T^{p}+c_{2}B^{q}+c_{3}\left(TB\right)^{r}\right)^{m}$ and upon enforcing the aforementioned conditions, the form is reduced to just $X=c_{1}T+c_{2}B$ where $c_{1},c_{2}\in\mathbb{R}$ and $c_{1}\neq-c_{2}$. One can easily see that a solution compatible with both $\Xi\equiv0$ and $\Upsilon\equiv0$ is $f(\lc{R})$.

\subsection{Branch \texorpdfstring{$\left\{ \Pi\protect\neq\text{const},\Upsilon\protect\neq0,\Xi\protect\neq0\right\} $}{}}

We will start with the most complex case that includes the full totally non-linear $f(T,B)$ models meaning those which will allow us to solve the constraint field in Eq.~(\ref{W=00005B0i=00005Dfinalsubaprox EXPANDED}) for $b$ as
\begin{equation}
    b = \frac{(a^{2}(12H^{3}f_{TT}+\dot{\Pi})+4Hk^{2}\Pi_{B})}{4aH^{2}k^{2}\Upsilon}\psi+\frac{(6a^{2}(\Pi_{B}\dot{H}+H^{2}f_{TT})-k^{2}\Pi_{B})}{2aHk^{2}\Upsilon}\phi\,,\label{bscalarsolution}
\end{equation}
which we replace into Eq.~(\ref{Wijfinalsubaprox}) in order to find
\begin{align}
    \frac{\psi}{\phi}= & \frac{2H\left(6a^{4}\dot{\Pi}(\Pi_{B}\dot{H}+H^{2}f_{TT})\right)}{-a^{4}\dot{\Pi}(12H^{3}f_{TT}+\dot{\Pi})+4a^{2}Hk^{2}(-2\Pi_{B}\dot{\Pi}+12H^{3}f_{BB}f_{TB}+H\Upsilon f_{T})-16H^{2}k^{4}\Xi}\nonumber \\
    & +\frac{2H\left(a^{2}k^{2}(-\Pi_{B}\dot{\Pi}-24H^{3}f_{BB}f_{TB}+2H\Upsilon f_{T}+24H\Xi\dot{H})-4Hk^{4}\Xi\right)}{-a^{4}\dot{\Pi}(12H^{3}f_{TT}+\dot{\Pi})+4a^{2}Hk^{2}(-2\Pi_{B}\dot{\Pi}+12H^{3}f_{BB}f_{TB}+H\Upsilon f_{T})-16H^{2}k^{4}\Xi}\,,\label{psiscalarsol1}
\end{align}
that we then substitute into Eqs.~(\ref{W00finalsubaprox}) so that we finally end up with
\begin{eqnarray}
    G_{\rm eff}&=&G\frac{A_{1}k^{2}+A_{2}k^{4}+A_{3}k^{6}}{A_{4}+A_{5}k^{2}+A_{6}k^{4}+A_{7}k^{6}}\label{GeffectivefTBbeta}\,,\\
    \Sigma&=&\frac{\Delta_{1}k^{2}+\Delta_{2}k^{4}+\Delta_{3}k^{6}+\Delta_{4}k^{8}+\Delta_{5}k^{10}}{\Delta_{6}+\Delta_{7}k^{2}+\Delta_{8}k^{4}+\Delta_{9}k^{6}+\Delta_{10}k^{8}+\Delta_{11}k^{10}}\,,\label{SIGMADEFLECTION}
\end{eqnarray}
where all the coefficients $A_{i}$ and $\Delta_{i}$ are presented in the Appendix \ref{sec:Geff-appendix}. One can further calculate the leading order terms of the above quantities by noticing that $A_{3}\propto\Xi$, $A_{7}\propto\Xi$ are the only coefficients, proportional to $\Xi$ and the same happens with the coefficients $\Delta_{5}\propto A_{3}$ and $\Delta_{11}\propto A_{7}$. This clarifies our choice for using $\Xi$ as an extra layer in branching. Hence the leading order parts read respectively
\begin{align}
    G_{\rm eff} & =G\frac{A_{3}}{A_{7}}=-G\frac{4\Upsilon}{36H^{2}(f_{BB}f_{TT}+2\Xi)+3\Upsilon f_{T}}\,.\label{GeffectivefTBbetaLEADING}
\end{align}
\begin{eqnarray}
    \Sigma & =- & \frac{\Upsilon}{\Upsilon f_{T}+12H^{2}\left(f_{BB}f_{TT}+2\Xi\right)}\,,
\end{eqnarray}
The models in this case assume the most possible general form they can from the class of $f(T,B)$, for example $f(T,B)=f_{1}(T)+f_{2}(T)f_{3}(B)+f_{4}(B)$.

\subsection{Branch \texorpdfstring{$\left\{ \Pi\protect\neq\text{const},\Upsilon\protect\neq0,\Xi=0\right\} $}{}}

A special case arises if $A_{3}=A_{7}\equiv0$ which means that $\Xi\equiv0$, giving that the leading order term for the gravitational effective constant is
\begin{eqnarray}
    G_{\rm eff} & =&G\frac{A_{2}}{A_{6}}\label{Geff =00039E=00003D0}\,,
\end{eqnarray}
and for the deflection parameter, we get
\begin{eqnarray}
    \Sigma & = & \frac{\Delta_{4}}{\Delta_{10}}=-\frac{A_{2}}{A_{6}}=-\frac{-A_{3}(4H\Upsilon f_{T}-5\Pi_{B}\dot{\Pi})}{4A_{7}\left(-(H\Upsilon f_{T}-2\Pi_{B}\dot{\Pi})+24H^{3}f_{BB}f_{TT}+12H^{3}f_{TB}(f_{TT}-\Upsilon)\right)}\,.\label{Sigma XI=00003D0}\
\end{eqnarray}
One can notice that $G_{\rm eff}$ becomes significantly more complicated since it depends on $A_2$ and $A_6$ (see~Appendix~\ref{sec:Geff-appendix}), and for that reason we explicitly calculate it for only two simple such models. The first one, is the popular $f(T)$ gravity models which up to next to leading order we find from Eq.~(\ref{Geff =00039E=00003D0}) 
\begin{eqnarray}
    G_{\rm eff} & = & \frac{a^{2}\dot{f}_{T}(12H^{3}f_{TT}+\dot{f}_{T})-4H^{2}k^{2}f_{T}f_{TT}}{4H^{2}f_{T}f_{TT}(6a^{2}H\dot{f}_{T}+k^{2}f_{T})}\,,\label{Geff f(T)-1}\\
    \Sigma & = & \frac{3a^{2}\dot{f}_{T}(8H^{3}f_{TT}+\dot{f}_{T})-8H^{2}k^{2}f_{T}f_{\text{TT}}}{2f_{T}\Big(a^{2}\dot{f}_{T}(12H^{3}f_{TT}-\dot{f}_{T})+4H^{2}k^{2}f_{T}f_{TT}\Big)}\,,
\end{eqnarray}
that correctly reproduce the usual leading order result $G_{\rm eff}=-G/f_{T}$ reported in Refs.\cite{Zheng:2010am,Nesseris:2013jea,Nunes:2016plz,Abedi:2018lkr}. The other, less known, model is Eq.~(\ref{f(T,B)=00003D-=0003B3T+f(B)}) for which (\ref{Geff =00039E=00003D0}) gives
\begin{eqnarray}
    G_{\rm eff} & =&G\frac{4H(2\dot{f}_{B}+H)}{(\dot{f}_{B}+2H){}^{2}}\,,\label{Geff -=0003B3T+f(B)}\\
    \Sigma & = & \frac{a^{2}\dot{f}_{B}{}^{2}+4Hk^{2}f_{BB}(2\dot{f}_{B}+H)}{3a^{2}f_{BB}\dot{f}_{B}\left(H^{2}(9\dot{f}_{B}+14H)-3\dot{H}(\dot{f}_{B}+2H)\right)+k^{2}f_{BB}(\dot{f}_{B}+2H)^{2}}\,.
\end{eqnarray}

\subsection{Branch \textmd{\normalsize{}\texorpdfstring{$\left\{ \Pi\protect\neq\text{const},\Sigma=0,\Xi=\Pi_{T}^{2}=\Pi_{B}^{2}\protect\neq0\right\} $}{}}}

In this branch $b$ completely drops out from Eq.~(\ref{W=00005B0i=00005Dfinalsubaprox EXPANDED}) and we can solve for $\psi$ as
\begin{equation}
    \frac{\psi}{\phi} = \frac{2H\Pi_{T}(k^{2}-6a^{2}\dot{H})}{4Hk^{2}\Pi_{T}-a^{2}\dot{\Pi}}\,,\label{psiscalarsolution}
\end{equation}
where we replace this solution into Eq.~(\ref{Wijfinalsubaprox}) and solve for $b$ as

\begin{equation}
    b = \frac{f_{T}\left(2H\Pi_{T}(6a^{2}\dot{H}+k^{2})-a^{2}\dot{\Pi}\right)+2\dot{\Pi}(k^{2}-6a^{2}\dot{H})(f_{TB}+\Pi_{T})}{(4Hk^{2}\Pi_{T}-a^{2}\dot{\Pi}){}^{2}}a\,\phi\,.\label{bscalarsolution2}
\end{equation}
Next we substitute both in Eq.~(\ref{W00finalsubaprox}) so that we can proceed and find $G_{\rm eff}$ as 
\begin{equation}
    G_{\rm eff} = G \frac{Z_{1}k^{2}+Z_{2}k^{4}+Z_{3}k^{6}}{Z_{4}+Z_{5}k^{2}+Z_{6}k^{4}+Z_{7}k^{6}}\,,\label{GeffectivefTBpsi}
\end{equation}
where again we omitted the rest of the cumbersome coefficients. The leading order contribution is then
\begin{equation}
    G_{\rm eff} = G\frac{Z_{3}}{Z_{7}}=-G\frac{4}{3(f_{T}+12H^{2}f_{TB})}
\end{equation}
In the same manner, we also calculate the deflection parameter 
\begin{equation}
    \Sigma = \frac{Y_{1}k^{2}+Y_{2}k^{4}+Y_{3}k^{6}+Y_{4}k^{8}}{Y_{5}+Y_{6}k^{2}+Y_{7}k^{4}+Y_{8}k^{6}+Y_{9}k^{8}}\,,
\end{equation}
where and to leading order 
\begin{equation}
    \Sigma = \frac{Y_{4}}{Y_{9}}=-\frac{1}{f_{T}+12H^{2}f_{TB}}
\end{equation}
which is a much simpler form than (\ref{Sigma XI=00003D0}).

\subsection{Branch \textmd{\normalsize{}\texorpdfstring{$\left\{ \Pi = {\rm const},\Sigma=0,\Xi=\Pi_{T}^{2}=\Pi_{B}^{2}\equiv0\right\} $}{}}}

The condition $\Pi_{T}^{2}=\Pi_{B}^{2}\equiv0$ means exactly that $\Pi=f_{T}+f_{B}\equiv c$ which is the condition to obtain $f(\lc{R})$ gravity (while not precisely $f(\lc{R})$ gravity when $c \neq 0$, it is dynamically equivalent). This is a pivotal branch because it is the only one where the antisymmetric part of the field equations is trivialised $W_{[0i]}\equiv0$ and also b completely drops out from the field equations. We solve $W_{ij}$ for $\psi$
\begin{eqnarray}
    \frac{\psi}{\phi} & = & \frac{a^{2}(12f_{RR}\dot{H}+f_{R})-2k^{2}f_{RR}}{a^{2}f_{R}-4k^{2}f_{RR}}\,,
\end{eqnarray}
next we substitute this in Eq.~(\ref{W00finalsubaprox}) so that we can proceed and find
as per usual to find
\begin{eqnarray}
    G_{\rm eff} & = & \frac{8k^{4}f_{RR}-2a^{2}k^{2}f_{R}}{-9a^{4}(f_{R}(4f_{RR}\dot{H}^{2}+H\dot{f}_{R})+4Hf_{RR}\dot{f}_{R}\dot{H})-2a^{2}k^{2}(-15Hf_{RR}\dot{f}_{R}+9f_{R}f_{RR}\dot{H}+f_{R}^{2})+6k^{4}f_{R}f_{RR}}\,,\\
    \Sigma & = & \frac{6k^{4}f_{RR}-2a^{2}k^{2}(6f_{RR}\dot{H}+f_{R})}{-9a^{4}(f_{R}(4f_{RR}\dot{H}^{2}+H\dot{f}_{R})+4Hf_{RR}\dot{f}_{R}\dot{H})-2a^{2}k^{2}(-15Hf_{RR}\dot{f}_{R}+9f_{R}f_{RR}\dot{H}+f_{R}^{2})+6k^{4}f_{R}f_{RR}}\,,
\end{eqnarray}
If one further employs the approximation $|\dot X| \sim H |X|$ where $X$ denotes background quantities, in conjunction with the matter dominated approximation $|f_R/(H^2f_{RR})| >> 0$ then  one will straightforwardly recover
\begin{eqnarray}
G_{{\rm eff}} & \sim & G\left(\frac{4}{3f_{R}}+\frac{1}{3(-f_{R}+3\frac{k^{2}}{a^{2}}f_{RR})}\right),\\
\Sigma & \sim & \frac{1}{f_{R}}.
\end{eqnarray}
which are the typical $f(\lc{R})$ results \cite{Tsujikawa:2007gd,DeFelice:2010aj} for $G_{\rm eff}$ and $\Sigma$.

\section{Conclusion and outlook\label{conclu}}

In this paper we have developed the theory of cosmological perturbations for the $f(T,B)$ models. This is mandatory to assess either the theoretical or observational viability of the models. We confronted, via tensor perturbations, the $f(T,B)$ theory against the recent multimessenger measurements that indicated speed of light propagation of GWs and it was in total agreement. We also obtained the simple and vital stability condition $f_T<0$ both from the vector and tensor perturbations. As for the scalar perturbations, we only studied the sub-horizon limit in order to find the equations governing the linear formation of cosmological structures which are required to confront the theories with the available cosmological precision data, but these equations have been lacking in the literature.

Most importantly, our study revealed the subtle and rather complicated branching of solutions that occurs in the scalar sector of perturbations. In all the subclasses of models in the different branches, we presented the post-Friedmannian parameters relevant for testing the models against the data, generalising in a non-trivial way the previously known results for $f(T)$ and $f(\lc{R})$ gravity. As a main result we highlight the Meszaros Eq.~(\ref{meszaroseq}), which can be now used for any particular $f(T,B)$ model to easily, by solving a simple homogeneous second order ODE, to solve the growth rate of cosmological structures at the observable scales. It is useful to note that this form of the equation suffices in all the different branches and thus e.g. a non-zero effective sound speed is not required unlike in, say Palatini $f(\lc{R})$ models. Thus the $f(T,B)$ class of models may in principle contain viable models which could, by modifying suitably the formation of structures, alleviate some of the tensions there exist in the $\Lambda$CDM interpretation of the data.

The detailed comparison of different models with the data is clearly outside of the scope of this paper. Arguably, a more urgent task is to clarify the number and nature of the degrees of freedom in the different branches of the $f(T,B)$ models. In the special case $f(T,B)=f(T)$ it is known that there are strongly coupled degrees of freedom in the cosmological background, and it will be interesting to see whether the models in the class (\ref{f(T,B)=00003D-=0003B3T+f(B)}) have this same problematical feature. We plan to clarify this issue in a forthcoming paper.

\begin{acknowledgments}
The authors would like to acknowledge networking support by the COST Action CA15117, CA16104 and CA18108. JLS would also like to acknowledge funding support from Cosmology@MALTA which is supported by the University of Malta. SB is supported by the European Regional Development Fund and the programme Mobilitas Pluss (Grant N$^\circ$ MOBJD423). JLS would like to acknowledge funding support from Cosmology@MALTA which is supported by the University of Malta. TSK was funded by the Estonian Research Council grants PRG356 “Gauge Gravity” and MOBTT86, and by the European Regional Development Fund CoE program TK133 “The Dark Side of the Universe”. V.G would like to thank J. Beltran and A. De Felice for useful and fruitful discussions. 
\end{acknowledgments}

\appendix

\section{Cosmological perturbations}

This section is devoted in presenting the most important quantities needed for the cosmological perturbations. 

\subsection{Background }

The non-zero components of the torsion tensor and superpotential, and the torsion and boundary term in the background (flat FLRW) are
\begin{eqnarray}
    T^{i}{}_{0j} & = & H\delta^{i}{}_{j}\,,\\
    S_{i}{}^{0j} & = & -H\delta^{i}{}_{j}\,,\\
    T & = & -6H^{2}\,,\\
    B & = & -6(3H^{2}+\dot{H})\,.
\end{eqnarray}
The matter content is fully conserved giving the standard conservation equation for a perfect fluid 
\begin{equation}
    \mathring{\nabla}_{\nu}\Theta_{\mu}{}^{\nu}:\qquad\dot{\rho}+3(\rho+P)=0\,.
\end{equation}

\subsection{Tensor perturbations}

The non-zero components of the torsion tensor and the superpotential are 
\begin{eqnarray}
\delta T^{i}{}_{0j} & = & \frac{1}{2}\dot{h}_{ij}\,,\\
\delta T^{i}{}_{jk} & = & \frac{1}{2}(\partial_{j}h_{ik}-\partial_{k}h_{ij})\,,\\
\delta S_{0}{}^{0i} & = & 0\,,\\
\delta S_{i}{}^{0j} & = & \frac{1}{4}\dot{h}_{ij}\,,\\
\delta S_{i}{}^{jk} & = & -\frac{1}{4a^{2}}(\partial_{j}h_{ik}-\partial_{k}h_{ij})\,,
\end{eqnarray}
while the scalars are 
\begin{equation}
\delta T=0\,,\quad\delta B=0\,.
\end{equation}

\subsection{Vector and pseudovector perturbations}

The non-zero components of the vectorial and pseudo vectorial perturbations for the torsion tensor and the superpotential are 
\begin{eqnarray}
\delta T^{0}{}_{0i} & = & a\dot{\beta}_{i}\,,\\
\delta T^{i}{}_{0j} & = & 2\partial_{i}\dot{h}_{j}-\frac{1}{a}\partial_{j}b_{i}-\epsilon_{kij}\dot{\sigma}_{k}\,,\\
\delta T^{0}{}_{ij} & = & a(\partial_{i}\beta_{j}-\partial_{j}\beta_{i})\,,\\
\delta T^{i}{}_{jk} & = & 2(\partial_{i}\partial_{j}h_{k}-\partial_{i}\partial_{k}h_{j})+\epsilon_{ijl}\partial_{k}\sigma_{l}-\epsilon_{ikl}\partial_{j}\sigma_{l}\,,\\
\delta S_{0}{}^{0i} & = & -\frac{1}{2a^{2}}\Big[2aH(b_{i}-\beta_{i})+\epsilon_{ilk}\partial_{k}\sigma_{l}\Big]\,,\\
\delta S_{i}{}^{0j} & = & -\frac{1}{2a}\Big[\frac{1}{2}\Big(\partial_{i}(b_{j}+\beta_{j}-a\dot{h}_{j})+\partial_{j}(b_{i}-\beta_{i}-a\dot{h}_{i})\Big)\Big]\,,\\
\delta S_{0}{}^{ij} & = & -\frac{1}{4a^{3}}\Big[\partial_{i}(b_{j}-\beta_{j}+2a\dot{h}_{j})-\partial_{j}(b_{i}-\beta_{i}+2a\dot{h}_{i})-2a\epsilon_{lij}\dot{\sigma}_{l}\Big]\,,\\
\delta S_{i}{}^{jk} & = & -\frac{1}{2a^{2}}\Big[\delta_{im}\epsilon_{kjl}\partial_{l}\sigma_{m}+\delta_{ij}\Big(2aH(b_{k}-\beta_{k})-a\dot{\beta}_{k}-2\partial^{2}h_{k}\Big)\nonumber \\
 & & -\delta_{ik}\Big(2aH(b_{j}-\beta_{j})-a\dot{\beta}_{j}-2\partial^{2}h_{j}\Big)-2\delta_{il}\partial_{k}\partial_{l}h_{j}+2\delta_{kl}\partial_{i}\partial_{j}h_{l}\Big]\,,
\end{eqnarray}
and the perturbations related to the torsion and boundary term scalars are 
\begin{eqnarray}
\delta T & = & 0\,,\\
\delta B & = & 0\,.
\end{eqnarray}

\subsection{Scalar and pseudo scalar perturbations}
\label{scalarappendix}

The components of the torsion tensor and the superpotential for scalar and pseudo scalar perturbations up to first order are 
\begin{eqnarray}
\delta T^{0}{}_{0i} & = & \partial_{i}(a\dot{\beta}-\phi)\,,\\
\delta T^{i}{}_{0j} & = & \partial_{i}\partial_{j}(\dot{h}-a^{-1}b)-\epsilon_{lij}\partial_{l}\dot{\sigma}-\dot{\psi}\delta_{ij}\,,\\
\delta T^{0}{}_{ij} & = & 0\,,\\
\delta T^{i}{}_{jk} & = & \delta_{ij}\partial_{k}\psi-\delta_{ik}\partial_{j}\psi+\delta_{il}(\epsilon_{klm}\partial_{j}\partial_{m}\sigma-\epsilon_{jlm}\partial_{k}\partial_{m}\sigma)\,,\\
\delta S_{0}{}^{0i} & = & -\frac{H}{a}\partial_{i}\Big(b-\beta-(aH)^{-1}\psi\Big)\,,\\
\delta S_{i}{}^{0j} & = & \Big[(2H\phi+\dot{\psi})\delta_{ij}+\frac{1}{2}\partial_{i}\partial_{j}(\dot{h}-a^{-1}b)-\frac{1}{2}\partial^{2}(\dot{h}-a^{-1}b)\delta_{ij}\Big]\,,\\
\delta S_{0}{}^{ij} & = & \frac{1}{2a^{2}}\epsilon_{ijk}\partial_{k}\dot{\sigma}\,,\\
\delta S_{i}{}^{jk} & = & \frac{1}{2a^{2}}\Big[\delta_{ik}\partial_{j}\Big(2aH(b-\beta)+\phi-\psi-a\dot{\beta}\Big)-\delta_{ij}\partial_{k}\Big(2aH(b-\beta)+\phi-\psi-a\dot{\beta}\Big)\Big]\,,
\end{eqnarray}
and the perturbations up to first order to the scalar torsion and boundary term become 
\begin{eqnarray}
\delta T & = & 4H\Big(3H\phi+3\dot{\psi}+\frac{1}{a}\partial^{2}b-\partial^{2}\dot{h}\Big)\,,\\
\delta B & = & -\Big[H\left(\frac{1}{a}\partial^{2}(6\beta-10b)-6(6\dot{\psi}+\dot{\phi}-2\partial^{2}\dot{h}+6H\phi)\right)+\frac{2}{a}\partial^{2}(\dot{\beta}-\dot{b})+\frac{2}{a^{2}}\partial^{2}(2\psi-\phi)\nonumber \\
 & & +2(\partial^{2}\ddot{h}-6\dot{H}\phi-3\ddot{\psi})\Big]\,.
\end{eqnarray}
Then, the perturbation conservation equations become 
\begin{eqnarray}
\mathring{\nabla}_{\mu}\Theta_{0}{}^{\mu} & = & \delta\dot{\rho}+3H(\delta P+\delta\rho)+\frac{\partial^{2}v(P+\rho)}{a}-3\dot{\psi}(P+\rho)+\partial^{2}\dot{h}(P+\rho)=0\,,\\
\mathring{\nabla}_{\mu}\Theta_{i}{}^{\mu} & = & \partial_{i}\Big[\delta P+(\rho+P)\Big(4aH(b+v-\beta)+\phi+a(\dot{b}-\dot{\beta}+\dot{v})\Big)+a(\dot{\rho}+\dot{P})(v+b-\beta)\Big]=0\,.
\end{eqnarray}

\subsection{Sub-horizon limit in the Newtonian gauge}

\begin{eqnarray}
\delta T & \simeq & -\frac{4H}{a}\left(k^{2}b-3aH(\psi+\phi)\right)\\
\delta B & \simeq & -\frac{2k^{2}}{a^{2}}(2abH-2\psi+\phi)\\
\delta f_{T} & \simeq & -\frac{2k^{2}}{a^{2}}\left(2abH\left(f_{TB}+f_{TT}\right)+f_{TB}(\phi-2\psi)\right)\label{deltafT}\\
\delta f_{B} & \simeq & -\frac{2k^{2}}{a^{2}}\left(2abH(f_{BB}+f_{TB})+f_{BB}(\phi-2\psi)\right)\,.\label{deltafB}
\end{eqnarray}

\section{\texorpdfstring{$G_{{\rm eff}}$}{} Calculations}\label{sec:Geff-appendix}

\subsection{Branch A}

\begin{align}
A_{1} & =-a^{4}\Upsilon\dot{\Pi}(\dot{\Pi}+12H^{3}f_{TT})\,,\\
A_{2} & =-4a^{2}H\Upsilon(2\Pi_{B}\dot{\Pi}-12H^{3}f_{BB}f_{TB}-H\Upsilon f_{T})\,,\\
A_{3} & =-16H^{2}\Xi\Upsilon\,,\\
A_{4} & =-3a^{6}\dot{\Pi}(\Pi_{B}+\Pi_{T})\Biggl[\\
 & \Pi'(6\dot{H}^{2}(\Pi_{B}-f_{TB})+6H^{2}\dot{H}(3f_{TB}-\Pi_{B})+18H^{4}(\Pi_{T}-f_{TB})-H^{2}f_{T}-H\dot{f}_{T})\\
 & +6H^{2}\biggl[\dot{H}(-24H^{3}(f_{TB}-\Pi_{B})(f_{TB}-\Pi_{T})+\Pi_{B}\dot{f}_{T})+H\dot{\Pi}^{2}\\
 & -12H\dot{H}^{2}(f_{TB}(\Pi_{B}-f_{TB})+\Xi)+H^{2}(2Hf_{T}+\dot{f}_{T})(f_{TB}-\Pi_{T})\biggl]\Biggl]\\
A_{7} & =12H^{2}\Xi(12H^{2}(f_{BB}f_{TT}+2\Xi)+\Upsilon f_{T})\,,\\
\nonumber \\
\Delta_{1}= & -a^{4}A_{1}\dot{\Pi}(12H\Pi_{B}\dot{H}-\dot{\Pi})\,,\\
\Delta_{2}= & a^{2}\left(-2H\Pi_{B}\dot{\Pi}(6a^{2}A_{2}\dot{H}-5A_{1})+a^{2}A_{2}\dot{\Pi}^{2}+8A_{1}H^{2}(-\Upsilon f_{T}-6\Xi\dot{H})\right),\\
\Delta_{3}= & a^{4}A_{3}\dot{\Pi}^{2}-2a^{2}H\Pi_{B}\dot{\Pi}(6a^{2}A_{3}\dot{H}-5A_{2})+8H^{2}\left(a^{2}A_{2}(-\Upsilon f_{T}-6\Xi\dot{H})+3A_{1}\Xi\right)\,,\\
\Delta_{4}= & 2H\left(4H\left(a^{2}A_{3}(-\Upsilon f_{T}-6\Xi\dot{H})+3A_{2}\Xi\right)+5a^{2}A_{3}\Pi_{B}\dot{\Pi}\right)\,,\\
\Delta_{5}= & 24A_{3}H^{2}\Xi\,,\\
\Delta_{6}= & 2a^{4}A_{4}\dot{\Pi}(\dot{\Pi}+12H^{3}f_{TT})\,,\\
\Delta_{7}= & 2a^{2}\left(\dot{\Pi}\left(a^{2}A_{5}(\dot{\Pi}+12H^{3}f_{TT})+8A_{4}H\Pi_{B}\right)\right)\\
 & +96A_{4}H^{4}(f_{BB}f_{TT}+\Xi)+48A_{4}H^{4}f_{TB}(f_{TT}-\Upsilon)-4A_{4}H^{2}\Upsilon f_{T}\,,\\
\Delta_{8}= & -2a^{4}A_{6}\dot{\Pi}(-\dot{\Pi}-12H^{3}f_{TT})+8a^{2}A_{5}H\\
 & +\left(-(H\Upsilon f_{T}-2\Pi_{B}\dot{\Pi})+24H^{3}(f_{BB}f_{TT}+\Xi)+12H^{3}f_{TB}(f_{TT}-\Upsilon)\right)+32A_{4}H^{2}\Xi\,,\\
\Delta_{9}= & 2a^{4}A_{7}-\dot{\Pi}(-\dot{\Pi}-12H^{3}f_{TT})+32A_{5}H^{2}\Xi\\
 & +8a^{2}A_{6}H\left(-(H\Upsilon f_{T}-2\Pi_{B}\dot{\Pi})+24H^{3}(f_{BB}f_{TT}+\Xi)+12H^{3}f_{TB}(f_{TT}-\Upsilon)\right)\,,\\
\Delta_{10}= & 8H\left(a^{2}A_{7}\left(-(H\Upsilon f_{T}-2\Pi_{B}\dot{\Pi})+24H^{3}(f_{BB}f_{TT}+\Xi)+12H^{3}f_{TB}(f_{TT}-\Upsilon)\right)+4A_{6}H\Xi\right)\,,\\
\Delta_{11}= & 32A_{7}H^{2}\Xi\,,
\end{align}

\subsection{Branch C}

\begin{align}
Z_{1}= & a^{4}\dot{\Pi}^{2}\,,\\
Z_{2}= & -8a^{2}H\dot{\Pi}\Pi_{T}\,,\\
Z_{3}= & 16H^{2}\Xi\,,\\
Z_{6}= & -a^{2}\dot{\Pi}^{2}(f_{TB}+\Pi_{T})\\
 & 12a^{2}H^{3}\Xi(5\dot{f}_{B}+72H^{3}f_{TT}+60a^{2}Hf_{TB}\dot{H})\nonumber \\
 & -12a^{2}H^{3}\dot{\Pi}\left(f_{TB}(2f_{TT}-7\Pi_{T})+2a^{2}f_{TT}\Pi_{T}\right)\nonumber \\
 & +4a^{2}Hf_{T}\Pi_{T}\left(-6H^{3}(f_{TT}+2\Pi_{T})+9a^{2}H\dot{H}\Pi_{T}+\dot{\Pi}\right)\,,\\
Z_{7}= & -12H^{2}\Xi(f_{T}+12H^{2}f_{TB})\,,\\
\nonumber \\
Y_{1}= & -a^{2}A_{1}(\dot{\Pi}+12H\dot{H}\Pi_{T})\,,\\
Y_{2}= & 6H\Pi_{T}(A_{1}-2a^{2}A_{2}\dot{H})-a^{2}A_{2}\dot{\Pi}\,,\\
Y_{3}= & 6H\Pi_{T}(A_{2}-2a^{2}A_{3}\dot{H})-a^{2}A_{3}\dot{\Pi}\,,\\
Y_{4}= & 6A_{3}H\Pi_{T}\,,\\
Y_{5}= & -2a^{2}A_{4}\dot{\Pi}\,,\\
Y_{6}= & -2a^{2}A_{5}\dot{\Pi}+8A_{4}H\Pi_{T}\,,\\
Y_{7}= & -2a^{2}A_{6}\dot{\Pi}+8A_{5}H\Pi_{T}\,,\\
Y_{8}= & -2a^{2}A_{7}\dot{\Pi}+8A_{6}H\Pi_{T}\,,\\
Y_{9}= & 8A_{7}H\Pi_{T}\,,
\end{align}

\end{document}